\documentclass[twoside]{ilcws08}
\usepackage[latin1]{inputenc}
\usepackage[dvips]{graphicx,epsfig,color}
\usepackage{wrapfig,rotating}
\usepackage{amssymb,amsmath,array}

\begin{document}

\title{
Precision Measurements of SM Higgs Recoil Mass and Cross Section for $\sqrt{s}$ of 230GeV and 250GeV at ILC } 
\author{Hengne Li, Francois Richard, Roman Poeschl, Zhiqing Zhang
\vspace{.3cm}\\
Laboratoire de l'Acc\'el\'erateur Lin\'eaire,  \\
Universit\'e Paris-Sud 11 et IN2P3-CNRS, BP34, 91898 Orsay Cedex, France 
}

\maketitle

\begin{abstract}

Precision measurements of SM Higgs recoil mass and cross-section are performed with the Higgs-Strahlung process $e^+e^- \rightarrow Z^0h^0$, with $Z^0 \rightarrow \mu^+\mu^-$ and $Z^0 \rightarrow e^+e^-$ at  230 GeV and 250 GeV center of mass energies, presuming $\rm{M_{h}=120GeV}$, assuming the integrated luminosity is $500fb^{-1}$, based on full simulations of several detector models for the ILC. Results are given and compared in this article, of the two center of mass energies, and of several detector models. According to the studies, the Higgs mass can be determined with a statistical error of 39 MeV and 3.6\% statistical error for the cross-section.

\end{abstract}

\section{Introduction}

The Higgs-strahlung process $e^+e^-\rightarrow Z^0h^0$ offers an unique opportunity for a model independent precision measurement of the Higgs Boson mass by means of the recoil mass to the $\rm{Z^0}$, gives as $M^{2}_{h^0}=s+M^{2}_{Z^{0}}-2E_{Z^0}\sqrt{s}$, where $\rm{Z^0}$ decays to $e^+e^-$ ($e$ channel) or $\mu^+\mu^-$ ($\mu$ channel). At the same time, the Higgs production cross-section and therefore also the coupling strength at the $Z^0h^0$ vertex can be determined, $g^2 \propto \sigma = N/\mathcal{L}\varepsilon$.

This article presents results of the studies performed for two center of mass energies ($\sqrt{s}$), at 230 GeV and  250 GeV. The 230 GeV is the optimal $\sqrt{s}$ in terms of the resolution of Higgs Mass according to a previous study\cite{richard}, where the $e$ channel is analyzed using detector model $\rm{LDC01Sc}$\cite{mokka} (LDC1). The 250 GeV is the benchmark scenario as proposed in the ILD\cite{ild} LOI, where both the $e$ channel and $\mu$ channel are analyzed using detector models $\rm{LDCPrime\_02Sc}$\cite{mokka} (LDCP), $\rm{LDC01\_06Sc}$\cite{mokka} (LDC6) and $\rm{LDC\_GLD\_01Sc}$\cite{mokka} (LDCG).

\section{Experimental Remarks}

In this study, the Higgs mass is presumed to be 120 GeV, the luminosity is assumed to be $\rm{500\ fb^{-1}}$, and unpolarized beams are assumed. The events generator is PYTHIA\cite{pythia}, simulation is performed using MOKKA\cite{mokka}, and the reconstruction is performed using MarlinReco\cite{marlin} and PandoraPFA\cite{pfa}. Both of the two $\sqrt{s}$ analyses include Beamstrahlung, ISR and FSR. The Beamstrahlung spectrum is generated using GUINEA-PIG\cite{gp}, and the interface to PYTHIA is CALYPSO\cite{calypso}.

\begin{table}[h]

\centerline{\begin{tabular}{|l|c|c|l|c|}
\cline{1-2} \cline{4-5}
\multicolumn{2}{|c|}{\texttt{$\sqrt{s}=230GeV$ $e$ channel}} &&
\multicolumn{2}{c|}{\texttt{$\sqrt{s}=250GeV$ $e$ and $\mu$ channels}}\\
\cline{1-2} \cline{4-5}
\texttt{Reactions}           & \texttt{$\sigma$}    &&
\texttt{Reactions}           & \texttt{$\sigma$} \\
\cline{1-2} \cline{4-5}
\texttt{\boldmath $Z^0h^0\rightarrow e^+e^-X$}      &  \texttt{\boldmath $6.3$ fb} &&
\texttt{\boldmath $Z^0h^0\rightarrow e^+e^-X$}      &  \texttt{\boldmath $7.5$ fb} \\
\cline{1-2} \cline{4-5}
$e^+e^-(\gamma)$            & $5.96\times10^5$ fb &&
\texttt{\boldmath $Z^0h^0\rightarrow \mu^+\mu^-X$}      &  \texttt{\boldmath $7.5$ fb} \\
\cline{1-2} \cline{4-5}
$\tau^+\tau^- \rightarrow e^+e^-4\nu$ & $146$ fb &&
$Z^0Z^0 \rightarrow e^+e^-f\bar{f}$ & $78.7$ fb \\
\cline{1-2} \cline{4-5}
$W^+W^- \rightarrow e^+e^-2\nu$ & $181$ fb &&
$Z^0Z^0 \rightarrow \mu^+\mu^-f\bar{f}$ & $79.0$ fb \\
\cline{1-2} \cline{4-5}
$Z^0/\gamma^*Z^0/\gamma^* \rightarrow e^+e^-f\bar{f}$ & $113$ fb &
\multicolumn{3}{c}{}\\
\cline{1-2} 
\end{tabular}}
\caption{Signals (bold) and backgrounds, and their cross-sections. Note, for $\sqrt{s}$ 230GeV backgrounds, only detector acceptance pre-cut is applied, $|cos\theta_{e^\pm}|<0.98$. Assume $\mathcal{L}=500fb^{-1}$ and unpolarized beams.}
\label{tab:xsec}
\end{table}

The cross-sections for signals and backgrounds are shown in Tab. \ref{tab:xsec}. For $\sqrt{s}$ at 230GeV, since only detector acceptance cut is applied, the Bhabha scattering ($e^+e^-$) gives a cross-section five orders of magnitude larger than that of the signal. The four detector models under study are composed of: TPC, time projection chamber; VXD, 5 single or 3 double layers of vertex detector; SIT/SET, 2 cylindrical layers of silicon strips inside and outside the TPC; FTD, pixel silicon disks in the forward region; ECAL, SiW electromagnetic calorimeter; HCAL, scintillator hadronic calorimeter. The magnetic fields in tracking system are 4 Tesla for LDC1 and LDC6, $3.5$ Tesla for LDCP, and $3$ Tesla for LDCG. The momentum resolution ($\Delta P/P^2$) is about $\rm{5 \times 10^{-5}\ GeV^{-1}}$ for the four detector models for momentum within 10 to 100 GeV\cite{track}.

A cut based electron ID is developed for the analysis at $\sqrt{s}=230\ \rm{GeV}$. The efficiency is greater than 99.5\%, and the rejection rate is 100\% for $\mu$, and larger than 98\% for $\pi$, for momentum within 20 to 80GeV, which covers the momentum range of the $e^+/e^-$ candidates of the signal. For the $\sqrt{s}$ at 250GeV analyses, MC information is employed temporarily for the lepton ID.

The final state lepton pair selection is based on a $\chi^2$ criteria that the invariant mass is the closest to the $Z^0$ mass.

\section{Background Rejection}

\begin{wrapfigure}{r}{0.5\columnwidth}
\vspace{-50pt}
\centering
\begin{minipage}[b]{0.25\columnwidth}
   \centering
   \includegraphics[width=0.99\columnwidth, origin=tl]{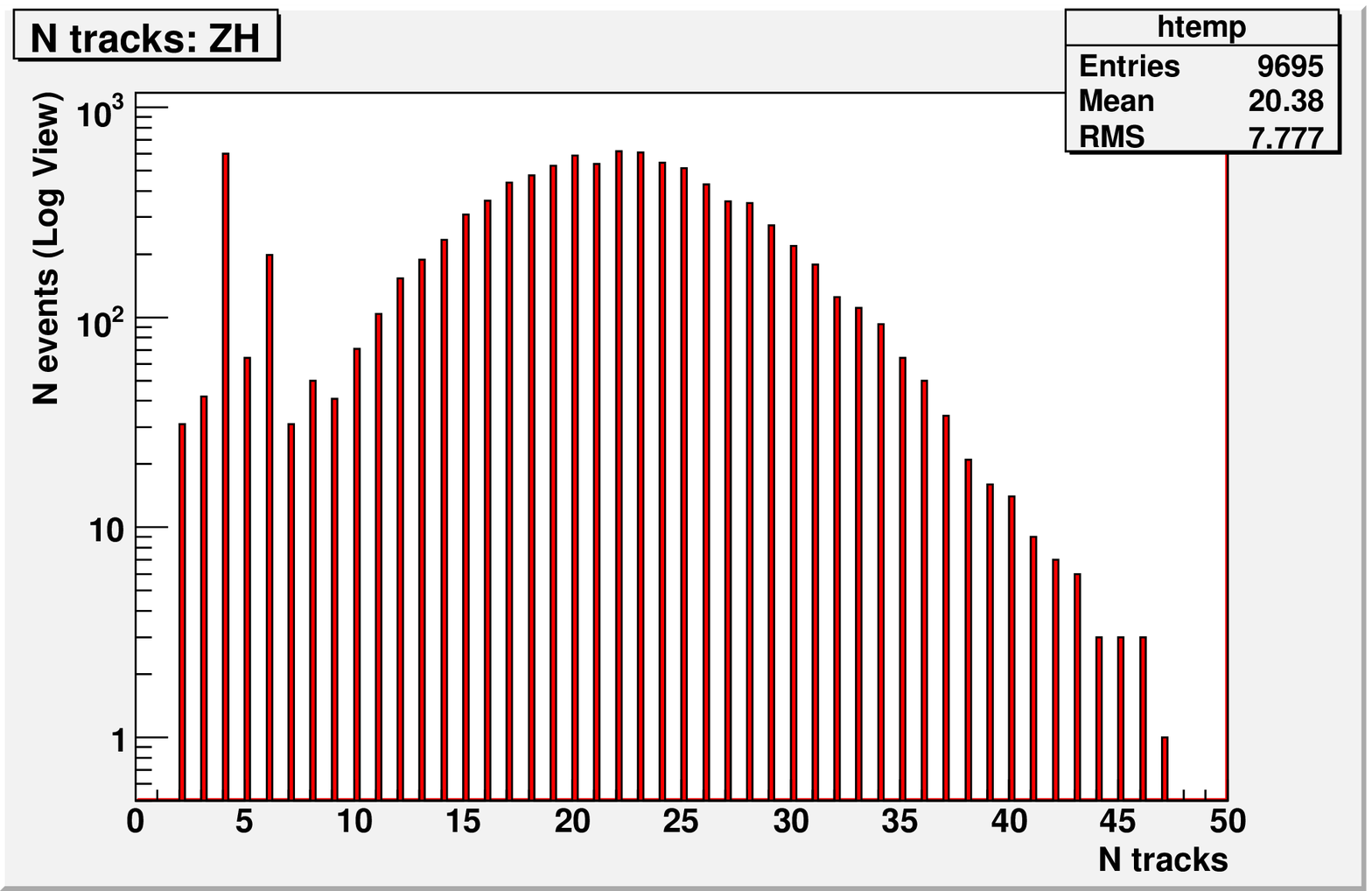}\\
\end{minipage}%
\begin{minipage}[b]{0.25\columnwidth}
   \centering
   \includegraphics[width=0.99\columnwidth, origin=tr]{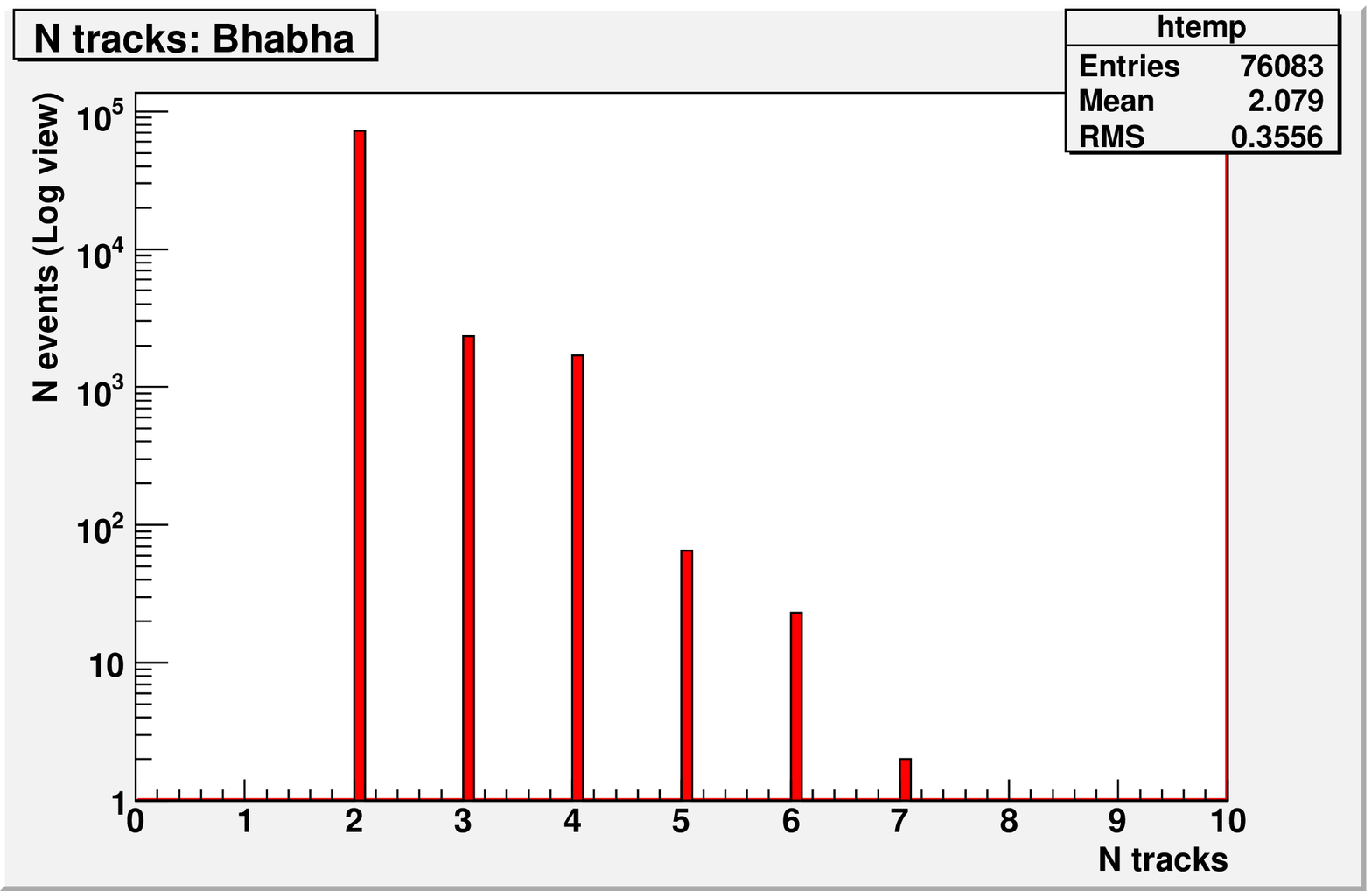}
\end{minipage}
\begin{minipage}[b]{0.25\columnwidth}
   \centering
   \includegraphics[width=0.99\columnwidth, origin=tr]{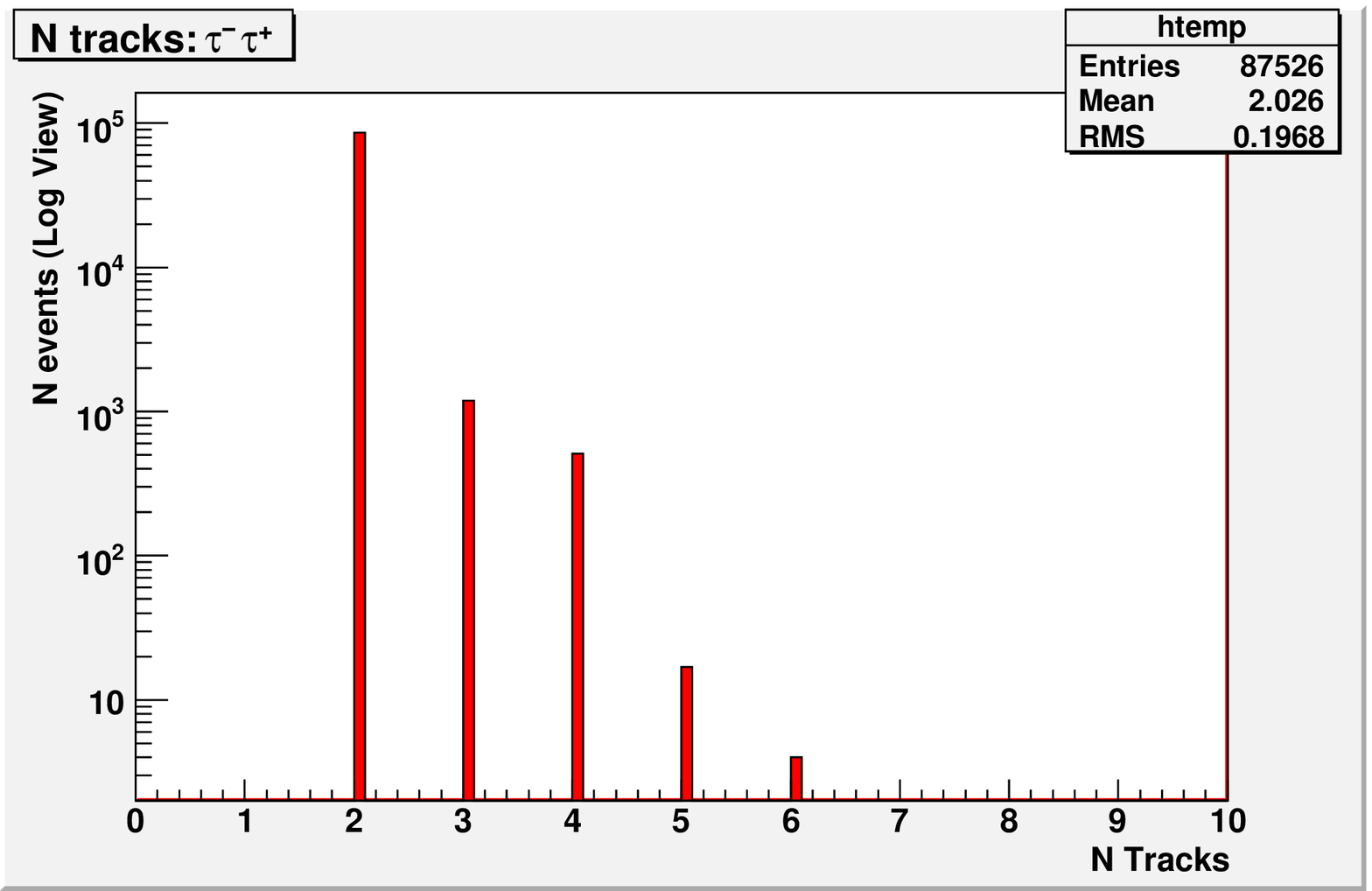}
\end{minipage}%
\begin{minipage}[b]{0.25\columnwidth}
   \centering
   \includegraphics[width=0.99\columnwidth, origin=tr]{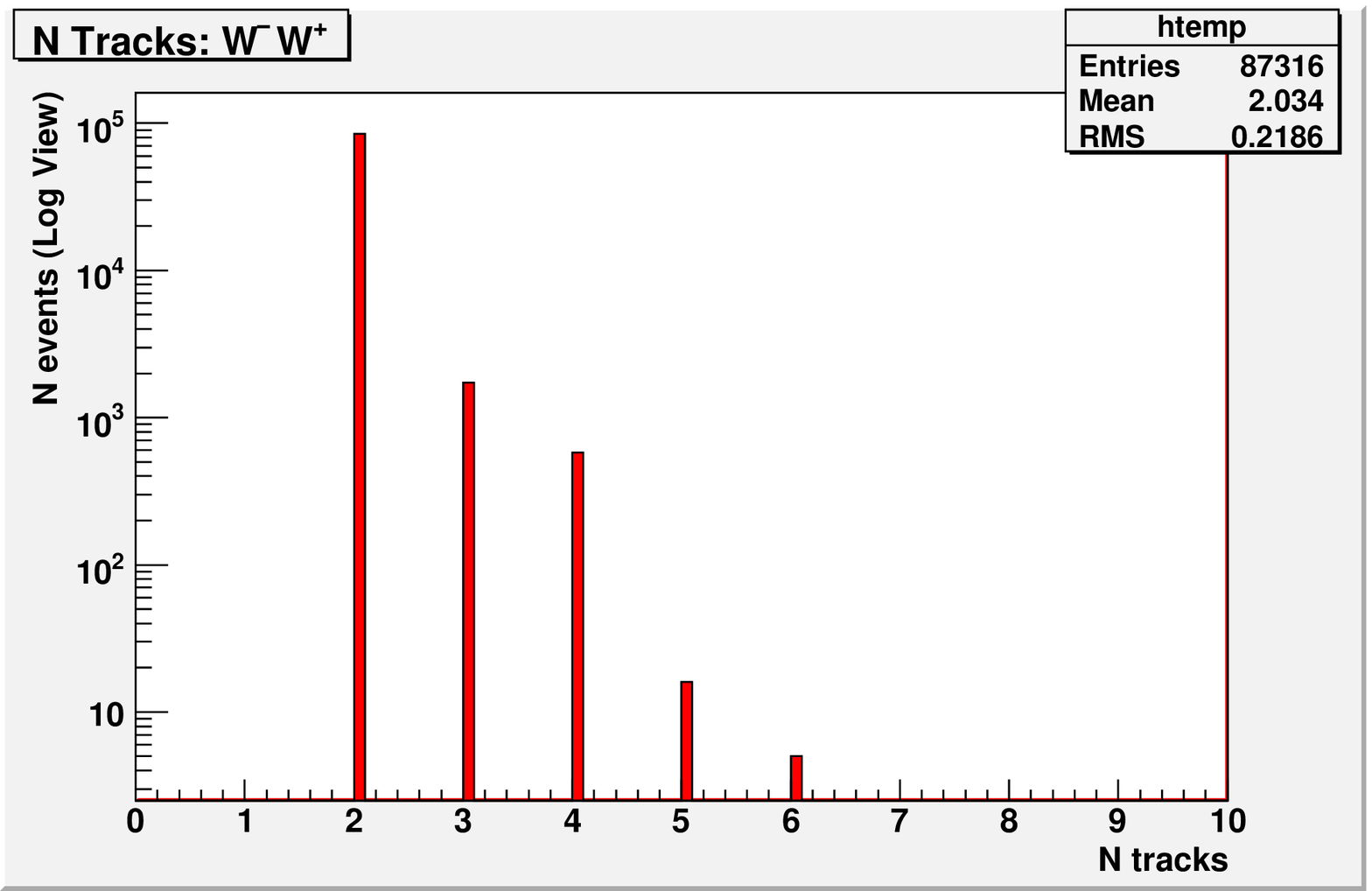}
\end{minipage}

\caption{$N_{tracks}$ distributions of: $ZH$ top-left; $Bhabha$ top-right; $\tau\tau$ bottom-left; $WW$ bottom-right.}\label{Fig:ntk}
\end{wrapfigure}

Standard Model (SM) Higgs with $\rm{M_h}$ $\sim$ 120 GeV dominantly decays to 2 jets\cite{higgs}. Together with $e^+e^-$ decayed from $Z^0$, more than $95\%$ of the signal have a multiplicity of larger than 4 in the final states. For the backgrounds $Bhabha$, $\tau\tau$ and $WW$, in their visible final states, the multiplicity is $2$. Due to bremsstrahlung, one electron may lead to several reconstructed tracks. Therefore, for the $\sqrt{s}$ at 230 GeV $e$ channel analysis, cut on $N_{tracks}>6$ is applied to reject all the three backgrounds totally, with a penalty of $\sim 8\%$ reduction of the signal. The $N_{tracks}$ distributions of signal and the three backgrounds are shown on Fig \ref{Fig:ntk}.

A Likelihood method is applied for the rejection of $ZZ$ background for both center of mass energies.

\begin{figure}[h]

\centering

\begin{minipage}[b]{0.3\columnwidth}
   \centering
   \includegraphics[width=0.99\columnwidth]{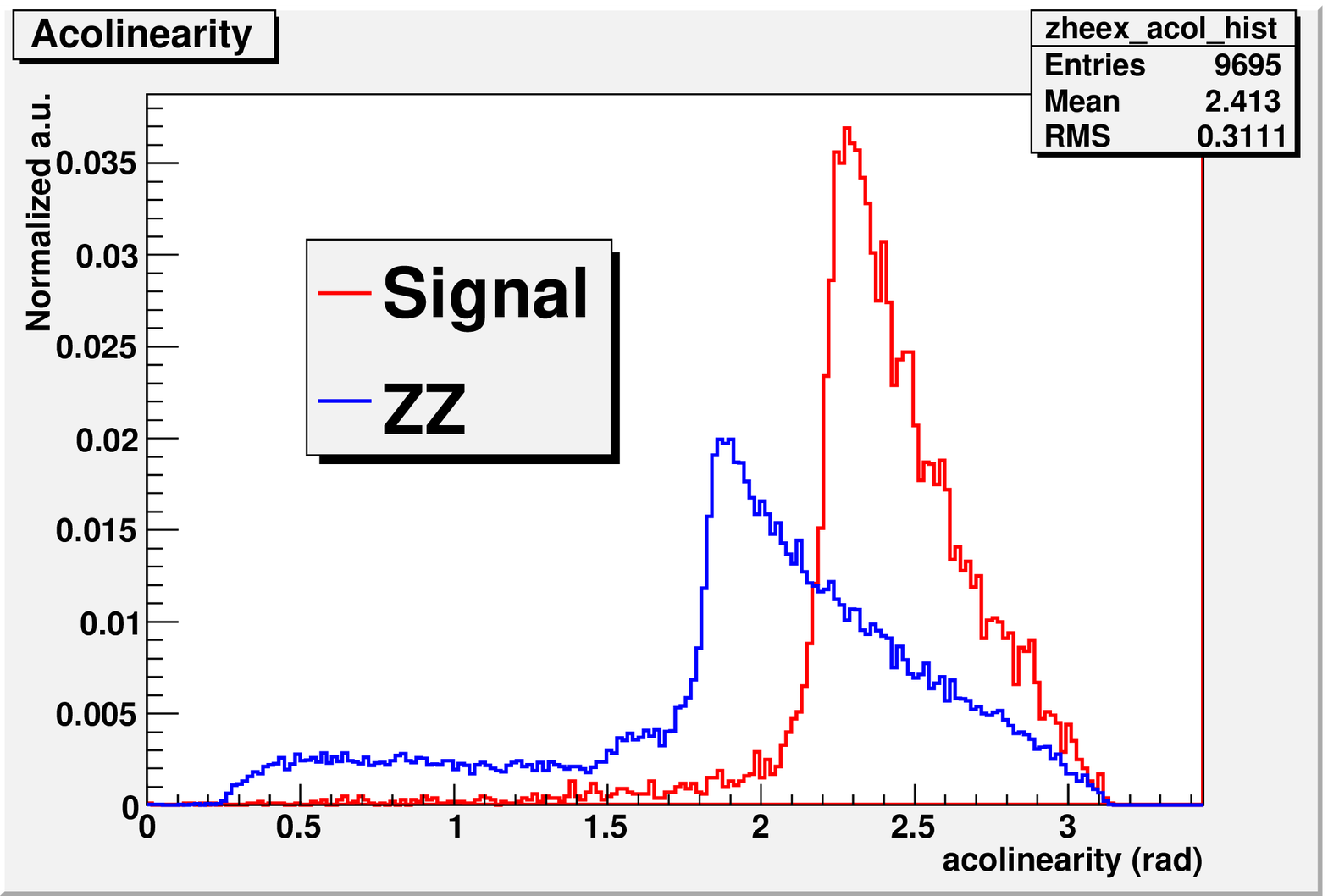}
\end{minipage}
\begin{minipage}[b]{0.3\columnwidth}
   \centering
   \includegraphics[width=0.99\columnwidth]{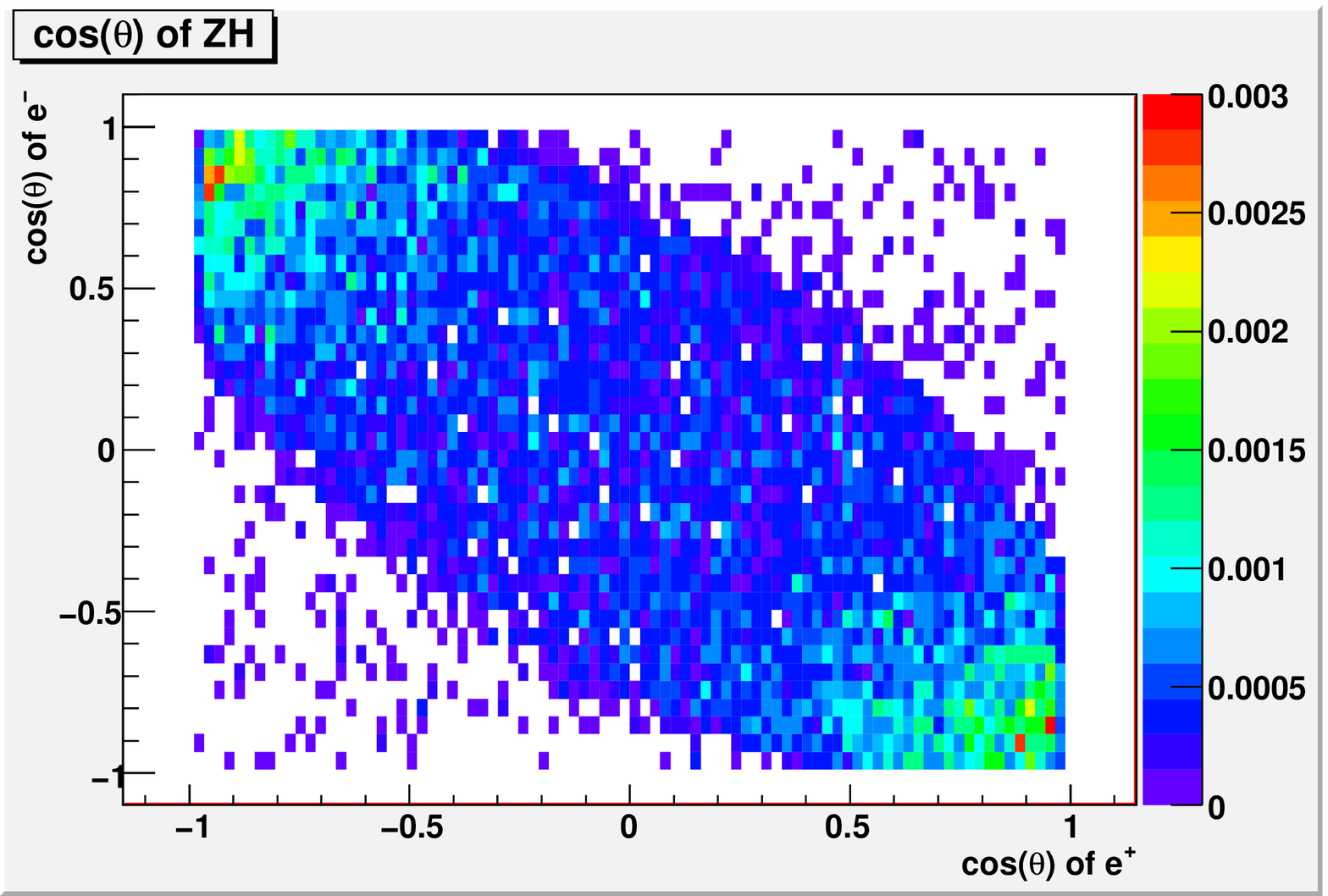}
\end{minipage}
\begin{minipage}[b]{0.3\columnwidth}
   \centering
   \includegraphics[width=0.99\columnwidth]{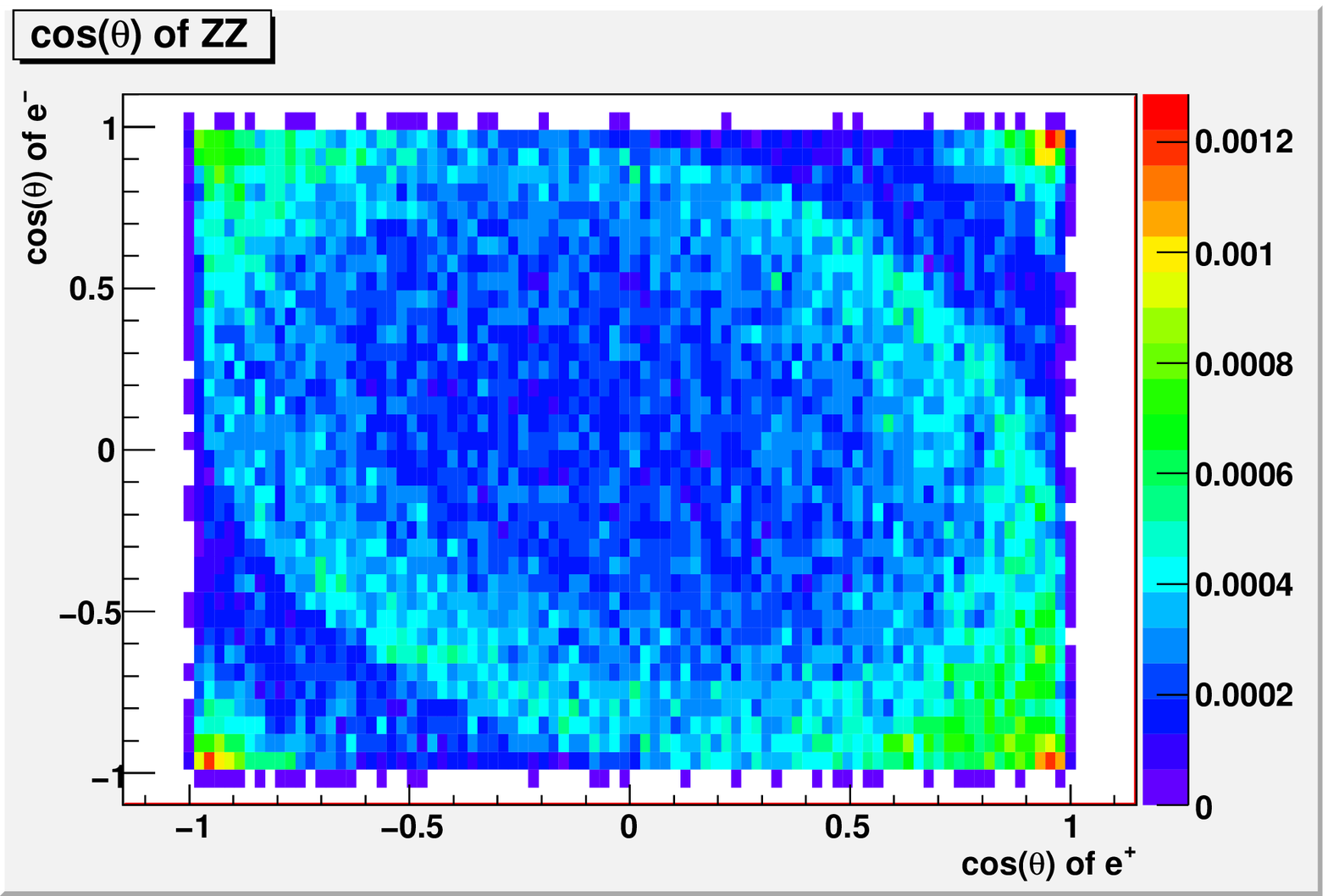}
\end{minipage}
\caption{PDFs of acolli\-ne\-ari\-ty (left) and $cos\theta_{e^-}$ vs. $cos\theta_{e^+}$ of signal (center) and $ZZ$ (right)}
\vspace{-6pt}
\label{Fig:pdf}

\end{figure}

The Likelihood of an event to be the signal is defined as $L_S=\prod{P^S_i}$, where the $P^S_i$ is the probability of the event to be the signal according to the PDF of the signal of the $i$th selection variable. Similarly, the Likelihood of an event to be the background is defined as $L_B=\prod{P^B_i}$. Thereafter, the Likelihood Fraction is defined as $f_{L}=L_S/(L_S+L_B)$, which is within $(0,1)$.

\begin{wrapfigure}{r}{0.56\columnwidth}
\centering
\begin{minipage}[b]{0.27\columnwidth}
   \centering
   \includegraphics[width=0.999\columnwidth, origin=tl]{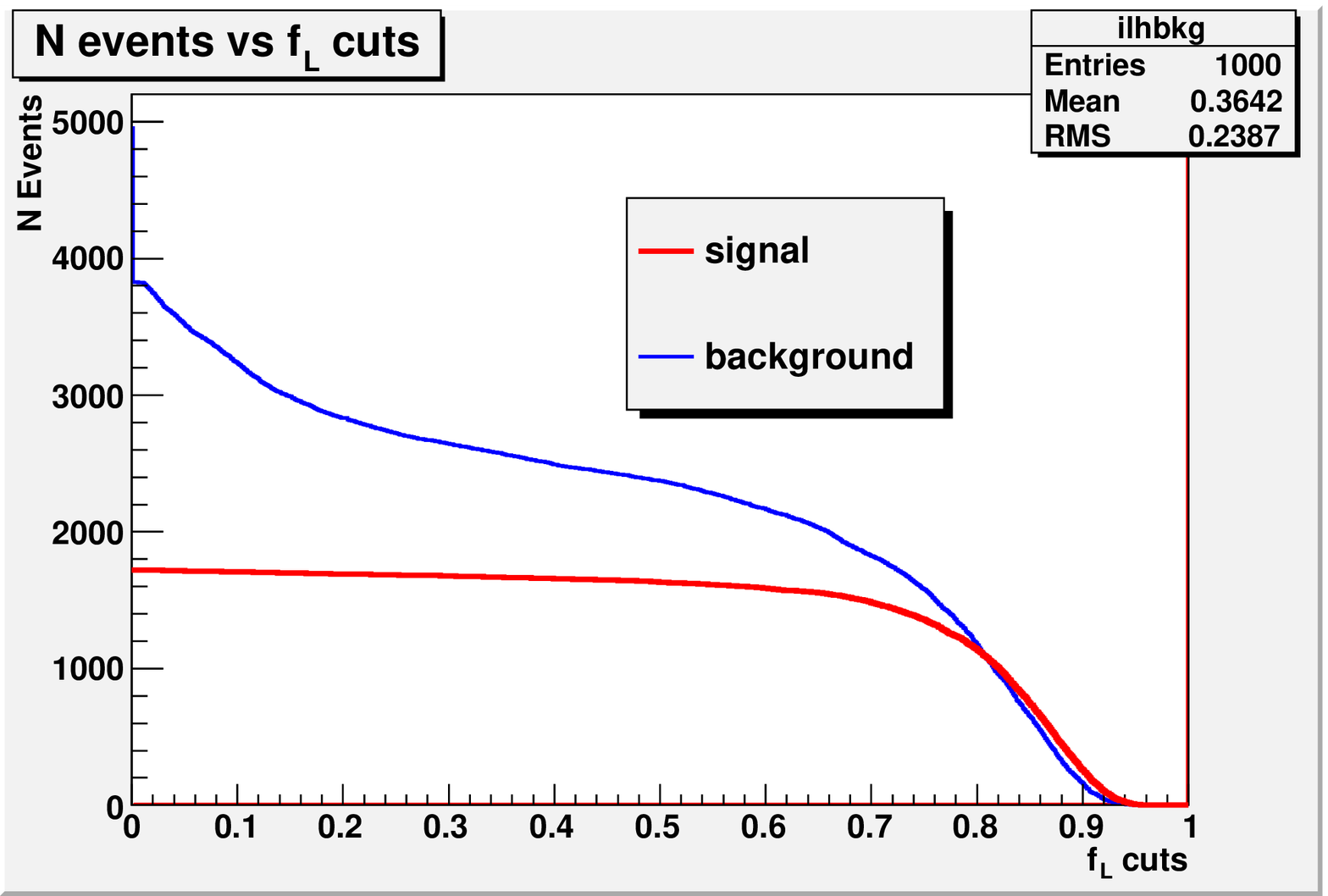}
\end{minipage}
\begin{minipage}[b]{0.27\columnwidth}
   \centering
   \includegraphics[width=0.999\columnwidth, origin=tr]{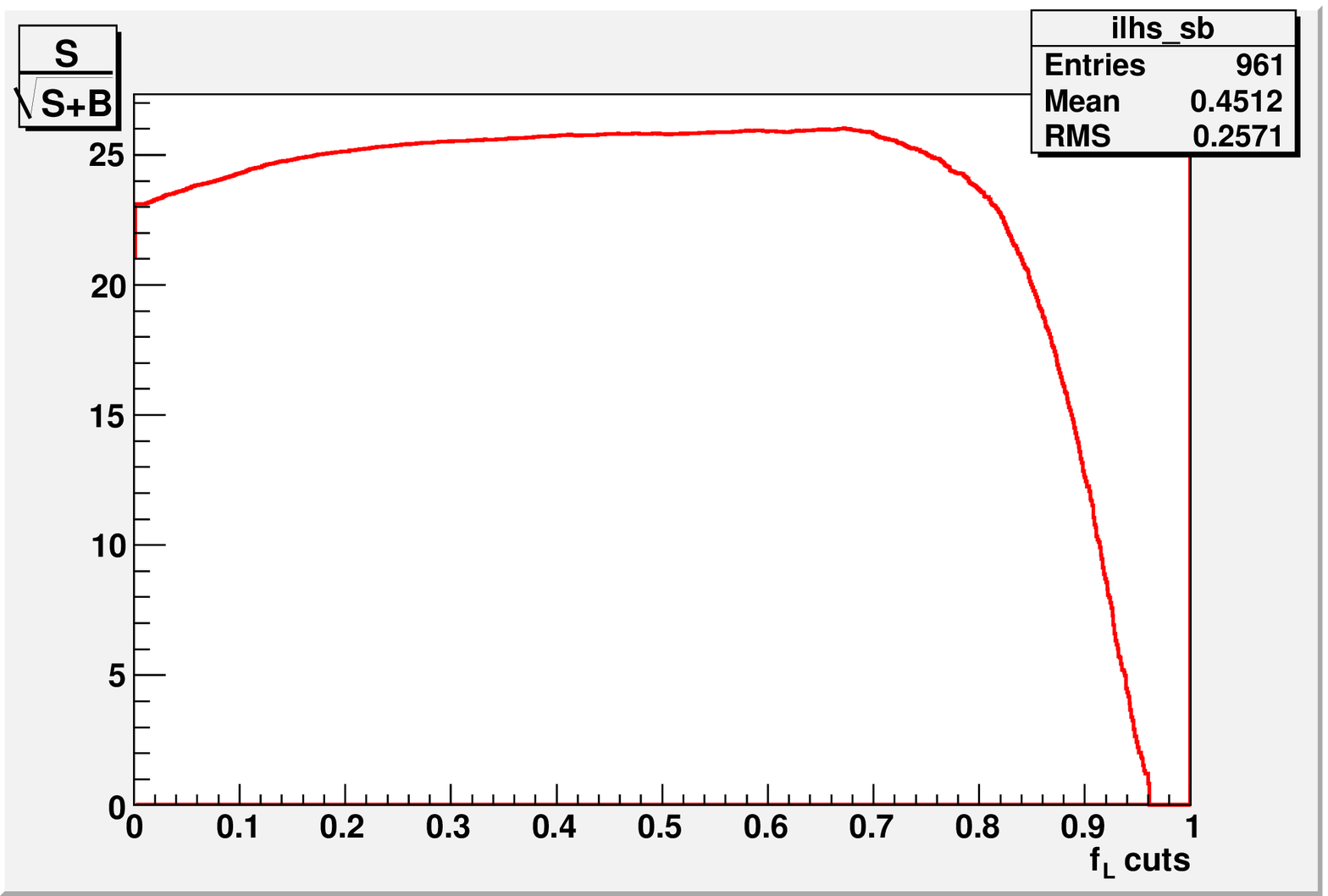}
\end{minipage}
\caption{Number of signal and number of backgrounds vs. $f_L$ cut. (top), and $S/\sqrt{S+B}$ vs. $f_L$ cuts (bottom).}
\vspace{-6pt}
\label{Fig:ssb}
\end{wrapfigure}

In order to be less dependent on the physics presumptions and to obtain a flat background distribution after its rejection, in this study, only two angular variables are employed: acol\-li\-ne\-a\-rity and $cos\theta_{e^-}$ vs. $cos\theta_{e^+}$. The PDFs of these two variables of $\sqrt{s}$ 230 GeV $e$ channel are shown in Fig. \ref{Fig:pdf}.

\begin{wraptable}{r}{0.5\columnwidth}
\centerline{\begin{tabular}{|c|c|c|c|}
\hline
Detectors & LDCP & LDC6 & LDCG\\
\hline
$e^+e^-X$ & $39.5\%$ & $37.2\%$ & $39.5\%$ \\
\hline
$\mu^+\mu^-X$ & $55.5\%$ & $52.7\%$ & $35.5\%$ \\
\hline
\end{tabular}}
\caption{Efficiencies of signal selection of $\sqrt{s}$ 250 GeV $e$ and $\mu$ channels, determined within $M_h$ window $\rm{118-135\ GeV}$.}
\vspace{-6pt}
\label{tab:ssb}
\end{wraptable}

The number of signal and background vs. $f_L$ cuts, and the significance $S/\sqrt{S+B}$ vs. $f_L$ cuts are shown in Fig. \ref{Fig:ssb}, which have the $N_{tracks}>6$ cut applied already, and they are determined within the fitting range of $M_h$ from 118 to 135 GeV.

According to the maximum of the significance, the $f_L$ cut is determined to be $f_L>6.7$, resulting in an overall signal selection efficiency of $55.5\%$ and a fraction of remained $ZZ$ background of $3.5\%$ for $\sqrt{s}$ at 230 GeV analysis. For the $\sqrt{s}$ at 250 GeV analysis, since only $ZZ$ background is considered, the same Likelihood method is applied for the rejection of $ZZ$ as that of the $\sqrt{s}$ at 230 GeV. The resulted efficiencies are shown in Tab. \ref{tab:ssb}.

\section{Fitting Methods}

\begin{wrapfigure}{r}{0.36\columnwidth}
\centering
\includegraphics[width=0.36\columnwidth]{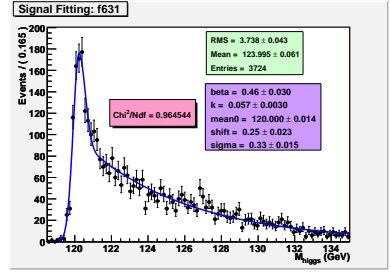}
\caption{Example fitting using the GCET formula.}
\label{Fig:sgf}
\end{wrapfigure}

Referred to fitting methods in previous studies\cite{jc}\cite{pwa}\cite{wolf}\cite{martin}\cite{manqi},  the \emph{Gaussian core for the Peak with Exponential complement for the Tail} (GPET) method is chosen and improved in this study, which eventually has both it-self and its first derivative continuous at all points, as shown in Eq.\ref{eq:fit}. It is a partial function, the left part is a pure Gaussian, the right part is a sum of Gaussian and Exponential with the fractions of contribution to be $\beta$ and $1-\beta$ respectively, and a factor $k$ is introduced in order to keep the maximum ($x_0$) been covered by the pure Gaussian. Fig.\ref{Fig:sgf} showed an example fitting to the signal using this formula with a $\chi^2/Ndf \sim 1$.

\begin{equation}
\begin{array}{ll}

f(x)=N \left\{ \begin{array}{ll} e^{-\frac{(x-x_0)^2}{2\sigma^2}} &: \frac{x-x_0}{\sigma} \le k \\ 
\beta e^{-\frac{(x-x_0)^2}{2\sigma^2}}+(1-\beta) e^{-(x-x_0) \frac{k}{\sigma}} e^{\frac{k^2}{2}} &: \frac{x-x_0}{\sigma} > k \end{array} \right.

\end{array}
\label{eq:fit}
\end{equation}

The GPET fit provides an explicit description of the final spectrum including the detector response. Together with the Higgs mass, the mass resolution can also be measured by this method.

However, due to the uncertainty of the detector response, the maximum of final recoil mass spectrum has a shift to value larger than the physics presumption ($M_h^{mc}=120\rm{\ GeV}$ in this study). A correction of the detector effects is required in order to restore to real mass. Since the shift ($\Delta{x}=x_0-M_h^{mc}$) comes from the detector effects, an explicit measurement of this shift from full detector simulation is possible and acceptable. 


For $\sqrt{s}$ at 230GeV analysis, beside $M_h^{mc}=120\rm{\ GeV}$, simulations and reconstructions for $M_h^{mc}=$  117, 118, 119, 121, 122 and 123 GeV are performed, too. The shifts ($\Delta{x}$) are determined subsequently. The mean of the shifts measured $\overline{\Delta{x}}=0.270GeV$ is taken as the correction, and the standard deviation $\sigma(\Delta{x})=0.021\rm{GeV}$, which is the uncertainty of the measurement of the shifts, is taken as the systematic error, for the Higgs mass measurement. For $\sqrt{s}$ 250GeV analyses, since no simulations are available for $M_h^{mc}$ besides 120GeV, the shifts measured from $M_h^{mc}=120\rm{GeV}$ simulations are taken as the corrections, while the statistical errors from the measurements of the shifts  are taken as the systematic errors. Actually, this systematic error can be eliminated by increasing the statistics of MC data samples in the determination of the shifts.

\begin{wrapfigure}{r}{0.56\columnwidth}
\centering
\includegraphics[width=0.56\columnwidth]{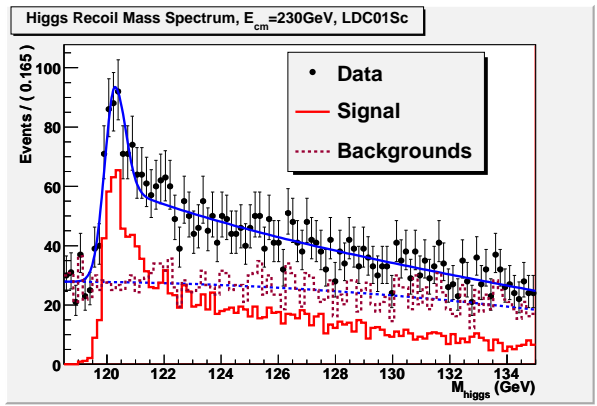}
\caption{The recoil mass spectrum of signal and backgrounds of $\sqrt{s}$ $230GeV$ $e$ channel.}
\label{Fig:ft230}
\vspace{-36pt}
\end{wrapfigure}

To describe the backgrounds spectrum, the Chebyshev Polynomial function with two coefficients is employed. 

\section{Results}

The final fitting to the signal with background is performed with the mass ($M_h=x_0-\Delta{x}$), the mass resolution ($\sigma$ in the pure Gaussian) and the number of signals as free parameters. A typical result is shown in Fig.\ref{Fig:ft230}, and the fitting results including the systematic errors of $M_h$ are listed in Tab.\ref{tab:rst}.

\begin{table}[h]
\vspace{-6pt}
\centering
\begin{tabular}{|c|c|c|c|c|c|}
\hline
 $\sqrt{s}$           &  Detector	& Chan-	& $M_h$ $(GeV)$        		& $\sigma$ $(fb)$     &   $\delta_m$ $(MeV)$     \\
 $(GeV)$ 		&  Model		& nel		& $(\pm stat.~err.\pm sys.~err.)$ &  $(\pm stat.~err.)$&  $(\pm stat.~err.)$\\
\hline
$230$		& LDC1& $e$	& $120.022\pm0.039\pm0.021$& $6.41\pm0.43(6.7\%)$ & $360\pm17$ \\

\hline
 			& LDCP	& $e$	& $119.973\pm0.047\pm0.039$& $7.82\pm0.52(6.6\%)$ & $540\pm25$ \\
\cline{3-6}
			& $$		& $\mu$	& $120.019\pm0.023\pm0.016$& $7.78\pm0.28(3.6\%)$ & $500\pm12$ \\
\cline{2-6}
$250$		& LDC6	& $e$	& $119.963\pm0.047\pm0.044$& $7.93\pm0.49(6.2\%)$ & $560\pm28$ \\
\cline{3-6}
			& $$	& $\mu$	& $119.994\pm0.023\pm0.016$& $7.45\pm0.27(3.6\%)$ & $550\pm12$ \\
\cline{2-6}
			& LDCG	& $e$	& $119.973\pm0.051\pm0.044$& $7.24\pm0.52(7.2\%)$ & $490\pm27$ \\
\cline{3-6}
			& $$		& $\mu$	& $120.003\pm0.029\pm0.020$& $7.45\pm0.32(4.3\%)$ & $530\pm15$ \\
\hline
\end{tabular}
\caption{Results of Higgs mass ($M_h$), cross-section ($\sigma$) and mass resolution ($\delta_m$). }
\label{tab:rst}
\end{table}

\section{Conclusion and Outlook}

According to the analysis results shown in Tab.\ref{tab:rst}, with the same luminosity, the statistical error of $\mu$ channel is about one half compared with the $e$ channel. This is due to the bremsstrahlung effect of electrons reduces the statistics on the maximum of the recoil mass. The results between different detector models are nearly the same, since the momentum resolutions are roughly the same\cite{track}. 

Improvements of this analysis can be expected with polarized beams, which may increase the cross-section of Higgs-strahlung reaction, and the left-handed polarized positron beam with right-handed polarized electron beam may suppress the WW background largely. For the rejection of Bhabha scattering and $ee\rightarrow \mu\mu$ in a model independent analysis, the $P_T$ of ISR photon can be used to balance the $P_T$ of the di-lepton system. Taken the Higgs mass to be 120 GeV, the $ZZ$ background can be further reduced by reconstruction both Z bosons.  Finally, the fitting range and fitting method may need to be optimized.


\begin{footnotesize}



%

\end{footnotesize}


\end{document}